\documentclass[useAMS,usenatbib,usegraphicx]{mn2e}
\usepackage{times}

\title[An HST/ACS View of the Inhomogeneous Outer Halo of M31]{An
HST/ACS View of the Inhomogeneous Outer Halo of M31\thanks{Based on
observations made with the NASA/ESA Hubble Space Telescope, obtained
at the Space Telescope Science Institute, which is operated by the
Association of Universities for Research in Astronomy, Inc., under
NASA contract NAS 5-26555.}}

\author[J.C. Richardson~et al.]
       {J.\,C.\,Richardson$^1$\thanks{jcr@roe.ac.uk}, A.\,M.\,N.\,Ferguson$^1$, A.\,D.\,Mackey$^1$,
       M.\,J.\,Irwin$^2$, S.\,C.\,Chapman$^2$,\newauthor 
       A.\,Huxor$^3$, R.\,A.\,Ibata$^4$, G.\,F.\,Lewis$^5$ ,
       N.\,R.\,Tanvir$^6$ \\ 
$^1$Institute for Astronomy, University of Edinburgh, Royal Observatory,
       Blackford Hill, Edinburgh EH9 3HJ, UK\\ 
$^2$Institute of
       Astronomy, University of Cambridge, Madingley Road, Cambridge,
       CB3 0HA, UK\\ 
$^3$Department of Physics, University of Bristol,
       Tyndall Avenue, Bristol, BS8 ITL, UK\\ 
$^4$Observatoire de
       Strasbourg, 11, rue de l'Universit\'{e}, F-67000 Strasbourg,
       France\\ 
$^5$ Sydney Institute for Astronomy, School of Physics, A29,
       University of Sydney, NSW 2006, Australia\\ 
$^6$Department of
       Physics and Astronomy, University of Leicester, LE1 7RH, UK\\ }

       \date{Accepted ?.  Received ?; in original form ?}

\pagerange{\pageref{firstpage}--\pageref{lastpage}}
\pubyear{2008}
  
\begin{document}

\maketitle

\label{firstpage}

\begin{abstract}

We present a high precision photometric view of the stellar
populations in the outer halo of M31, using data taken with the Hubble
Space Telescope Advanced Camera for Surveys (HST/ACS). We analyse the
field populations adjacent to 11 luminous globular clusters which
sample the galactocentric radial range $18\la ~{\rmn{R}} \la
100~{\rmn{kpc}}$ and reach a photometric depth of $\sim 2.5$
magnitudes below the horizontal branch (m$_{F814W}\sim 27$ mag). The
colour-magnitude diagrams (CMDs) are well populated out to $\sim
60~{\rmn{kpc}}$ and exhibit relatively metal-rich red giant branches,
with the densest fields also showing evidence for prominent red
clumps. We use the Dartmouth isochrones to construct metallicity
distribution functions (MDFs) which confirm the presence of dominant
populations with $<\![\mbox{Fe/H}]\!> \approx -0.6$ to $-1.0$~dex and
considerable metallicity dispersions of 0.2 to 0.3 dex (assuming a
10~Gyr population and scaled-Solar abundances). The average
metallicity over the range 30--60~kpc is [Fe/H]$=-0.8\pm 0.14$~dex,
with no evidence for a significant radial gradient. Metal-poor stars
([Fe/H]$\leq -1.3$) typically account for $\la 10-20\%$ of the
population in each field, irrespective of radius. Assuming our fields
are unbiased probes of the dominant stellar populations in these
parts, we find that the M31 outer halo remains considerably more
metal-rich than that of the Milky Way out to at least 60~kpc.

\end{abstract}


\section{Introduction}
\label{sec:intro}

Understanding the nature and origin of stars at extreme distances from
galaxy centres is of paramount importance. Modern theories predict
that the stellar halos of massive galaxies result from the continued
accretion and disruption of smaller satellite systems [from
\cite{SZ78} to \citealt{BJ05, deLucia08}]. The stellar populations
these systems donate to their host halos are unique in terms of their
mix of ages, chemical compositions and kinematics. In the outer halos
of galaxies, where the mixing timescales are of the order of several
gigayears \citep{JHB96}, fossil signatures of these accretion events
are expected to be well-preserved and detailed studies of the stellar
populations in these parts can constrain the number, nature and
timescales of both recent and ancient accretion events
\citep[e.g.,][]{Johnston08}. Such information is required in order to
rigorously test hierarchical growth models against the more
traditional monolithic collapse models, in which halo stars form
entirely {\it in situ} \citep[e.g.,][]{ELBS}.

The last decade has seen tremendous advances in our understanding of
our nearest large neighbour, M31. A particular highlight has been the
first detailed exploration of the large-scale stellar distribution and
content outside the main body of the galaxy. The Isaac Newton
Telescope wide-field imaging survey of M31 mapped a $\sim 100\times
100$~kpc$^2$ region down to $\approx3$ magnitudes below the red giant
branch (RGB) tip and revealed copious substructure in the spatial
distribution of stars \citep{Ibata01a, F02, Irwin05}. Follow-up work
with the Hubble Space Telescope Advanced Camera for Surveys (HST/ACS)
and the DEIMOS spectrograph on Keck has shown that almost all of this
inner halo substructure can be explained by a combination of giant
stream debris and perturbed disk material \citep{Ibata04, Ibata05,
F05, Brown06a, F07, Gilbert07, Faria07, Richardson08}, suggesting that
there has only been one significant accretion event onto the inner
halo of M31 over the last several gigayears.

Attention has recently been turned to the more remote regions of M31's
halo. Using MegaCam on the Canada-France-Hawaii Telescope (CFHT),
efforts are underway to map the stellar distribution out to a radius
of 150~kpc around M31, with an extension towards M33
\citep{I07,McConnachie08}. In addition to the discovery of several new
dwarf galaxies and halo globular clusters (GCs) \citep{Martin06,
Mackey06, Mackey07, Huxor08, McConnachie08}, this survey has also
uncovered new faint debris streams as well as an extremely extended
stellar halo with a shallow radial fall-off of $\Sigma_V(R) \propto
R^{-1.9}$. Spectroscopy of individual RGB stars in these outlying
regions has yielded the first evidence for a pressure-supported metal
poor ([Fe/H]$\approx -1.4$) population in the halo \citep{Chapman06,
Kal06b, Koch08}, although it remains unclear whether this component -
as opposed to a more metal-rich population associated with discrete
substructures - dominates the stellar halo mass in these parts.

In this paper, we present a high precision view of the stellar content
in the outer halo of M31 using data taken with the HST/ACS. We analyse
the field populations adjacent to 11 luminous GCs which lie at
projected radii of $18~{\rm kpc} \leq R \leq 100~{\rm kpc}$.  The
HST/ACS data yield accurate photometry with high completeness to
$\sim$2.5 magnitudes below the horizontal branch and thus enable the
construction of colour-magnitude diagrams (CMDs) and metallicity
distribution functions (MDFs) that are a vast improvement over those
obtained from ground, where Galactic foreground and background galaxy
contamination pose significant challenges at even moderately bright
magnitudes. 

The paper is outlined as follows. In \S~2 we describe the observations
and subsequent treatment of the data while in \S~3 we compare and
contrast the detailed CMDs of the fields and derive their metallicity
distribution functions. In \S~4 we discuss our results in the context
of theoretical predictions and current literature and in \S~5 we
summarise our findings.

\begin{figure}
  \centering
  \includegraphics[width=8.0 cm]{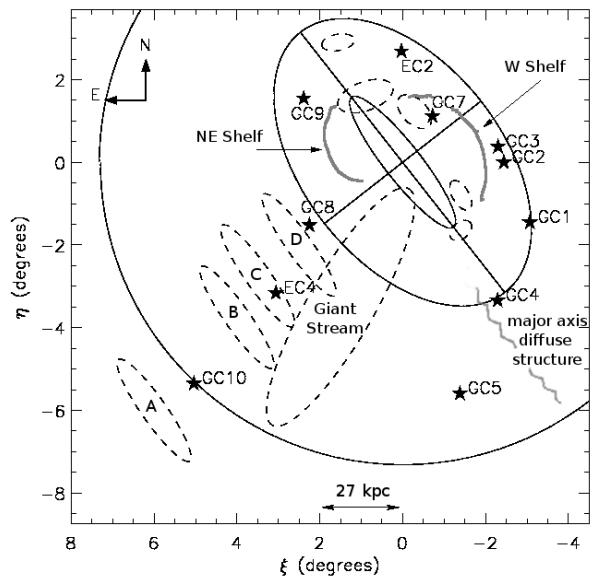}
  \caption{A schematic map of M31 with the globular cluster positions
    marked as black stars. The inner ellipse represents M31's main
    stellar disk ($i=77^{\circ}$ and R = $2^{\circ}$ or 27 kpc) while
    the outer ellipse has a radius of 55 kpc and is flattened to $b/a$ =
    0.6. The outer circle marks a radius of $100~{\rm kpc}$. Cartoons of
    the most significant outer halo substructures discovered to date are
    labelled. The inner halo substructures examined in
    \protect\cite{Richardson08} are highlighted with unlabelled dashed
    ellipses. Moving from the northeast to the southwest, they are the
    NE Clump, N Spur, NGC205 Loop, Claw and G1 Clump.}
  \label{fig:map}
\end{figure}


\section[]{Observations and data reduction}
\label{sec:obs}

\begin{table*}
 \begin{minipage}{140mm}
  \caption{Observational Information\label{tab:info}}
  \begin{tabular}{@{} c c c c c c c c c c @{} }
\hline Field & R.A. & Dec & $R$$^a$ & $E(B-V)$$^b$ & $R_{cut}$$^c$
& $F_{fg}$$^d$ & $<\![Fe/H]\!>$$^e$ & $F_{mp}$$^f$\\

  & (J2000.0) & (J2000.0) & (kpc) & & ($^{\prime\prime}$) &
 ($\%$) & (dex) & ($\%$) \\
\hline

    GC7F & 00:38:49.4 & 42:22:48.0 & 18.2 & 0.062 & 40 &  0.3 &
    -0.78 & 12.0\\
    GC3F & 00:30:27.3 & 41:36:20.4 & 31.8 & 0.075 & 45 &  2.7 &
    -0.82 & 17.0\\
    GC2F & 00:29:44.9 & 41:13:09.8 & 33.4 & 0.073 & 45 &  3.3 &
    -0.87 & 21.4\\
    EC2F & 00:42:55.1 & 43:57:28.5 & 36.8 & 0.087 & 50 &  1.0 &
    -0.81 & 15.5\\
    GC8F & 00:54:25.0 & 39:42:55.5 & 37.1 & 0.051 & 45 &  3.8 &
    -0.60 & 12.6\\
    GC9F & 00:55:44.0 & 42:46:16.1 & 38.9 & 0.099 & 45 &  1.6 &
    -0.67 & 10.0 \\
    GC1F & 00:26:47.8 & 39:44:45.5 & 46.4 & 0.070 & 40 &  10.3 &
    -0.67 & 10.3\\
    GC4F & 00:31:09.9 & 37:53:59.7 & 55.2 & 0.058 & 45 &  4.7 &
    -1.03 & 34.6\\
    EC4F & 00:58:15.5 & 38:03:01.1 & 60.0 & 0.049 & 40 &  2.5 &
    -0.78 & 15.5 \\
    GC5F & 00:35:59.7 & 35:41:03.6 & 78.5 & 0.065 & 45 & 44.0 &
    $\cdots$ & $\cdots$ \\
    GC10F& 01:07:26.4 & 35:46:49.7 & 99.9 & 0.055 & 45 & 16.8 &
    $\cdots$ & $\cdots$ \\
\hline
\end{tabular}
\medskip

$^a$Projected radial distance assumes $D_{M31} = 785~{\rm kpc}$
\citep{McC05}.\\ $^b$$E(B-V)$ values are interpolated from the
reddening map of \cite{Schlegel98}.\\ $^c$$R_{cut}$ is the radial cut
imposed to exclude GC stars.\\ $^d$$F_{fg}$ is the percentage of
foreground stars predicted in the region of the CMD used to compute
the MDF (through large-area realisations of the Besan{\c c}on model).\\
$^e$$<\![Fe/H]\!>$ is the average metallicity calculated from the MDF
(Section 3.1).\\ $^f$$F_{mp}$ is the metal-poor fraction, given by
relative number of stars in the MDF that have ([Fe/H]$\leq$-1.3~dex).

\end{minipage}
\end{table*}

The observations were obtained with the Wide Field Channel of the
HST/ACS as part of the Cycle 13 program GO 10394 (PI Tanvir).  The
original aim of this program was to obtain deep high resolution
photometry of outer halo GCs in M31 that were discovered in the course
of the the Isaac Newton Telescope and CFHT MegaCam surveys
\citep[][see]{Huxor05, Martin06, Huxor08} and results from this
analysis are reported elsewhere (\citealt{Mackey06, Mackey07}; Tanvir
et al. in prep.). In the present paper, we focus instead on the field
populations imaged alongside the GCs.

Fields were observed in the HST/ACS F606W and F814W filters for
$\sim1800{\rm s}$ and $\sim3000{\rm s}$ respectively, with small
dithers between various sub-exposures.  Full details of the
photometric reduction are provided in \cite{Mackey06}. In brief, we
used the ACS module of the DOLPHOT program \citep{Dolphin2000} to
obtain point-spread function (PSF) fitting photometry for each image.
Next, quality information provided by DOLPHOT was used to clean the
source catalogue of any non stellar sources such as background
galaxies and blended stars. Specifically, tests showed that strong
stellar detections satisfied the conditions $-0.5< \mbox{sharpness}
<0.9$, $-1.7< \mbox{roundness} <1.7$ and signal-to-noise ratio $>$
4. The data were further pruned according to the distribution of the
magnitude error, $\chi^2$ of the PSF, and sharpness as a function of
magnitude. Finally, magnitudes were transformed to the VEGAmag scale
using the zero-points of \cite{Sirianni05} and were corrected for
foreground reddening by interpolating within the maps of
\cite{Schlegel98}.

Thirteen GC fields were imaged in total. Two of these sample rich
substructure in the inner halo of M31 and have previously been
presented in \cite{Richardson08}; we will not discuss these fields
further here.  Extensive artificial star tests on those fields
demonstrated they were $>80\%$ complete at $m_{F606W,0}=27.3~{\rm
mag}$ and $m_{F814W,0}=27.0~{\rm mag}$. Given that they are a factor
of $\sim100$ times more crowded than fields analysed in this paper,
these values serve as a lower limit to our completeness. Stars
belonging to the GCs have been excluded from our analysis by masking a
circular region of radius $40-50^{\prime\prime}$ around each GC (see
Table~\ref{tab:info}). This corresponds to one tidal radius ($R_t$)
for GC10 and at least $2R_t$ for all other clusters (Tanvir et al. in
prep). Table~\ref{tab:info} lists the location and projected radius
from the centre of M31 for each pointing, as well as the reddening
value derived from the \cite{Schlegel98} maps. Field names follow the
nomenclature of \cite{Mackey06, Mackey07}. For example, GC1F is the
field population around the classical globular cluster GC1, and EC2F
is the field population associated with extended globular cluster EC2.

Figure 1 shows the locations of the GC fields around M31. All but one
of the fields probe what we will refer to as the outer halo ($R >
30~{\rm kpc}$) of M31. This boundary marks a key transition point in
the radial surface brightness profile of the galaxy where the
behaviour switches from a steep inner fall-off (often described with
an R$^{1/4}$ law) to a shallower power-law form (\citealt{Irwin05},
Ibata et al. 2007; hereafter I07). Using the extended disk parameters
of PA$=38.1^{\circ}$ and $i=64.7^{\circ}$ from Ibata et al. (2005),
the field locations analysed here correspond to de-projected radii of
$40-230{\rm~kpc}$ which makes it highly unlikely that they contain any
substantial disk component.


\begin{figure*}
\centering
\includegraphics[width=13.0 cm]{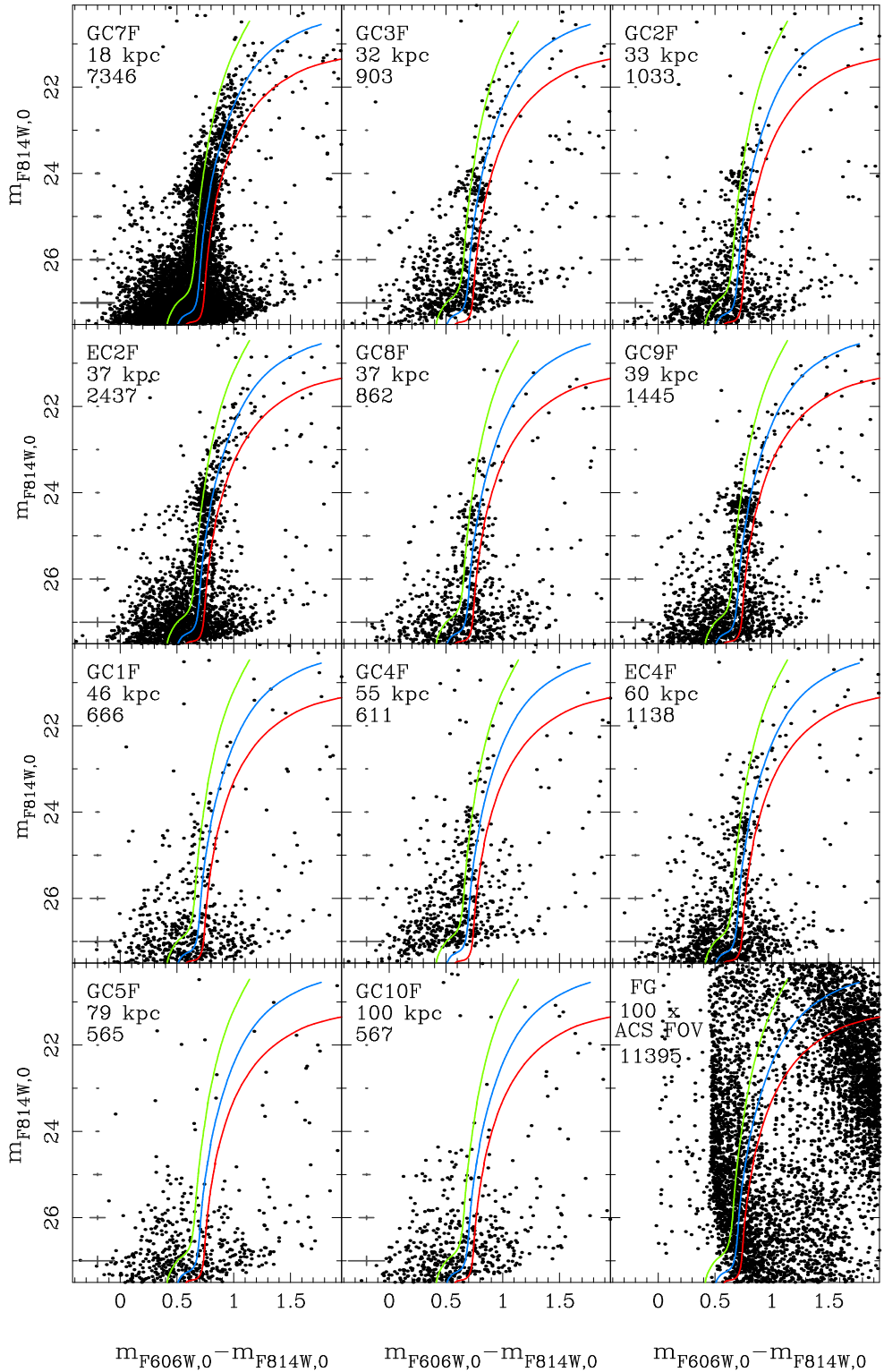}
  \caption{The extinction-corrected CMDs of the outer halo
  fields. Listed below the field name are the projected radial
  distance and the total number of sources on the CMD. Isochrones from
  the Dartmouth Stellar Evolution Database (DSED; \citealt{DSED}) with
  age 10 Gyr and [Fe/H]=-1.5 (green), [Fe/H]=-0.7 (blue) and
  [Fe/H]=-0.2 (red) have been plotted for comparison. They have been
  shifted to the distance of M31 (m-M)=24.47
  \citep{McC05}. Typical photometric errors (grey crosses) are
  plotted. GC5F and GC10F do not show convincing RGBs but show a
clustering of sources at faint magnitudes. Other fields have CMDs whose density at the faint
  end scales with the density of the red clump suggesting that some
  fraction of these sources are genuine stars. The last panel shows
  the MW foreground (FG) contamination predicted by the Besan{\c c}on
  Galactic model (\citealt{Besancon}) over an area 100 times larger
  than our ACS fields (centred on GC2F, a representative field). This
  large area realisation is included to produce a smooth
  representation of the foreground CMD morphology to compare with the
  GC fields.  }
  \label{fig:cmd}
\end{figure*}

\section{Analysis}
\label{sec:anal}
\subsection{Colour-Magnitude Diagrams (CMDs)}
\label{sec:cmdplots}

Figure~\ref{fig:cmd} shows the CMDs plotted in order of increasing
projected galactocentric radius. The average photometric errors
returned by DOLPHOT are indicated towards the left hand side of the
panels. Theoretical isochrones of [Fe/H]=$-1.5,-0.7,-0.2$~dex for
[$\alpha$/Fe]$=0$ and age $10~{\rm Gyr}$ have been obtained from the
Dartmouth Stellar Evolution Database (DSED; \citealt{DSED}) and are
superimposed. They have been shifted to the distance modulus of M31
($m-M=24.47$; \citealt{McC05}).

The CMDs range from being well-populated ($\sim 7000$ stars at R =
18~kpc) to being very sparse ($\leq 600$ stars at R $\geq 80$~kpc). As
a result, it is prudent to first consider the level of contamination
that might be present from the foreground Milky Way (MW)
population. Since our fields span almost 10 degrees in Galactic
latitude, the foreground population towards each field is expected to
vary substantially depending on the proximity of the line-of-sight to
the Galactic Plane. We have used the Besan\c{c}on Galactic model
\citep{Besancon} to obtain large area realisations (100 $\times$
larger than the ACS field of view) of the foreground population
towards each field which we subsequently scale down to the size of an
ACS pointing in order to reduce the shot noise in a given
simulation. The error function measured in each field was fed into the
model and the results were de-reddened in the same way as the data. We
find the number of foreground stars predicted to lie in the region of
the CMD used to derive the MDFs (\S~\ref{sec:mdf}) is less than or
equal to $10\%$ in all but the two outermost fields (see $F_{fg}$ in
Table 1). Furthermore, we find that the colour and magnitude
distribution of foreground stars is unlike that of our observed
CMDs. The bottom right-hand panel of Figure 2 shows the CMD of a large
area realisation of the Besan\c{c}on model towards a representative
field (GC2F) for increased resolution. This CMD contains two orders of
magnitude more stars than are predicted to fall within a single ACS
field and is intended as a comparison only. The CMD of the foreground
clearly does not reproduce any of the distinct features observed in
the GC fields, such as the red clump at $m_{F606W,0}-m_{F814W,0} =
0.8~{\rm mag}$, $m_{F814W,0} = 24.0~{\rm mag}$, nor does it follow the
curvature of the RGBs. We can therefore be confident that foreground
contamination is not a major issue for most of our fields.

Figure~\ref{fig:cmd} shows prominent RGB sequences in all fields out
to $R=60~{\rm kpc}$ with the bulk of the stars well-bounded by the
[Fe/H] = -1.5 dex and -0.2 dex isochrones. The mean RGB loci are
consistent with a moderately high metallicity of [Fe/H]$\sim -0.7$
although one field, GC4F, appears more metal poor with most stars
lying between the [Fe/H]$=-1.5$ and $-0.7$ dex isochrones. The widths
of the RGBs are considerably larger than the photometric error bars
for magnitudes brighter than $m_{F814W,0}\sim26.0~{\rm mag}$
suggesting an intrinsic spread in metallicity. It is worth considering
whether line-of-sight effects could artificially create this width. In
overlaying the isochrones, we have assumed that the stars in each
field all lie at the systemic distance of M31, whereas in reality they
will span a range of line-of-sight distances. We checked the effect a
distance spread would have on the CMDs by plotting isochrones with
varying distance moduli and measuring the resulting colour width. In
the extreme case of shifting the $-0.7$~dex isochrone to lie $\pm
60~{\rm kpc}$ in front (behind) M31, this induced a colour broadening
of $0.02$~mag at $m_{F814W,0} = 25.0$ mag and $0.05$~mag at
$m_{F814W,0} = 22.0$ mag. This accounts for less than $10\%$ and
$15\%$ of the width of the RGB at those magnitudes respectively and
implies the RGB spreads are for the most part intrinsic.

A red clump (RC) due to core helium burning stars is detected in most
fields with a well-populated RGB (GC7F, GC3F, GC2F, EC2F, GC9F, EC4F)
and indicates the presence of an intermediate age and/or metal-rich
population \citep{G&S01}. There is a tentative detection of an old,
metal-poor extended horizontal branch (HB) in a few fields (e.g. GC7F,
GC3F, EC2F, GC9F) which can be traced from $m_{F606W,0} - m_{F814W,0}
= 0.50$ mag, $m_{F814W,0} = 24.24$ mag to $m_{F606W,0} - m_{F814W,0} =
0.00$ mag, $m_{F814W,0} = 25.75$ mag. This feature may also be present
in GC8F and EC4F but there are too few stars detected to be certain.
Indeed, the sparseness of most of the CMDs makes it impossible to do
comparative analyses of the RC/HB morphologies in different fields, as
was done in \cite{Richardson08}.

Figure~\ref{fig:cmd} also highlights the presence of star count (i.e.
surface brightness) and stellar population inhomogeneities throughout
the halo. For example, EC2F, GC8F and GC9F lie in the projected radial
range of $\sim37-39~{\rm kpc}$ yet their total star counts fluctuate
by a factor of three ($2437 \pm 49, 862 \pm 29$ and $1445 \pm 38$
respectively) indicating genuine variations larger than the
Poisson uncertainties. Furthermore, EC4F and GC4F lie in the projected
radial interval $55-60~{\rm kpc}$, however EC4F exhibits a wider RGB
than GC4F. The former also contains twice as many stars ($1138 \pm
34$) as GC1F ($666 \pm 26$), a field which lies 15~kpc further
in. Since the fields have a uniform depth and completeness these
patterns indicate the that they do not probe a smooth well-mixed
stellar component of M31.

Although we detect stars in fields GC5F ($R = 79~{\rm kpc}$) and GC10F
($R = 100~{\rm kpc}$), the CMDs are too sparsely populated to reliably
constrain their stellar populations. Most of the sources in these CMDs
cluster around the base of the RGB sequences and while at least some
of them will be genuine RGB stars (note that in the more populated
fields, the density of stars with $m_{F814W,0} < 26.0$ scales with the
RGB and RC), there will also be a high contribution from unresolved
background galaxies and foreground stars. Without knowledge of the
true background populations along these sight-lines, it is impossible
to disentangle these different components at faint magnitudes
($m_{F814W,0} > 25.5$ mag).

\subsection{Metallicity Distribution Functions (MDFs)}
\label{sec:mdf}

\begin{figure}
\centering 
\includegraphics[width=6.0cm]{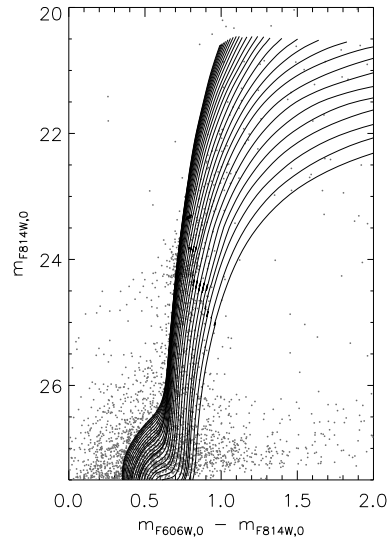}
  \caption{The CMD of EC2F (grey dots) on top of which is shown the
  theoretical RGB sequences of DSED \citep{DSED} used to compute the
  metallicity distributions. The isochrones have a Solar-scaled
  abundance mixture, age 10 Gyr and span the range [Fe/H] = $-2.5$ to
  $+0.3$ in 0.1 dex steps (left to right).  They have been shifted to
  (m-M) = 24.47.}  \label{fig:isocs}
\end{figure}

In order to further characterise the stellar populations in each
field, we determine the MDFs using a set of theoretical isochrones
from the DSED (\citealt{DSED}) with age 10~Gyr and a Solar-scaled
abundance mixture. Figure~\ref{fig:isocs} shows the RGB sequences of
the selected isochrones shifted to the distance of M31 and overlaid
onto the photometry of EC2F. We have selected isochrones which form a
finely-spaced grid ranging from [Fe/H]$=-2.5$ dex to $+0.3$ dex in
[Fe/H]$=0.1$ dex steps. With such sampling, the metallicities of RGB
stars can simply be derived by comparing their
$m_{F606W,0}-m_{F814W,0}$, $m_{F814W,0}$ values to the isochrone grid
and assigning the metallicity of the closest isochrone.

\begin{figure*}
\centering 
\includegraphics[width=16.0cm]{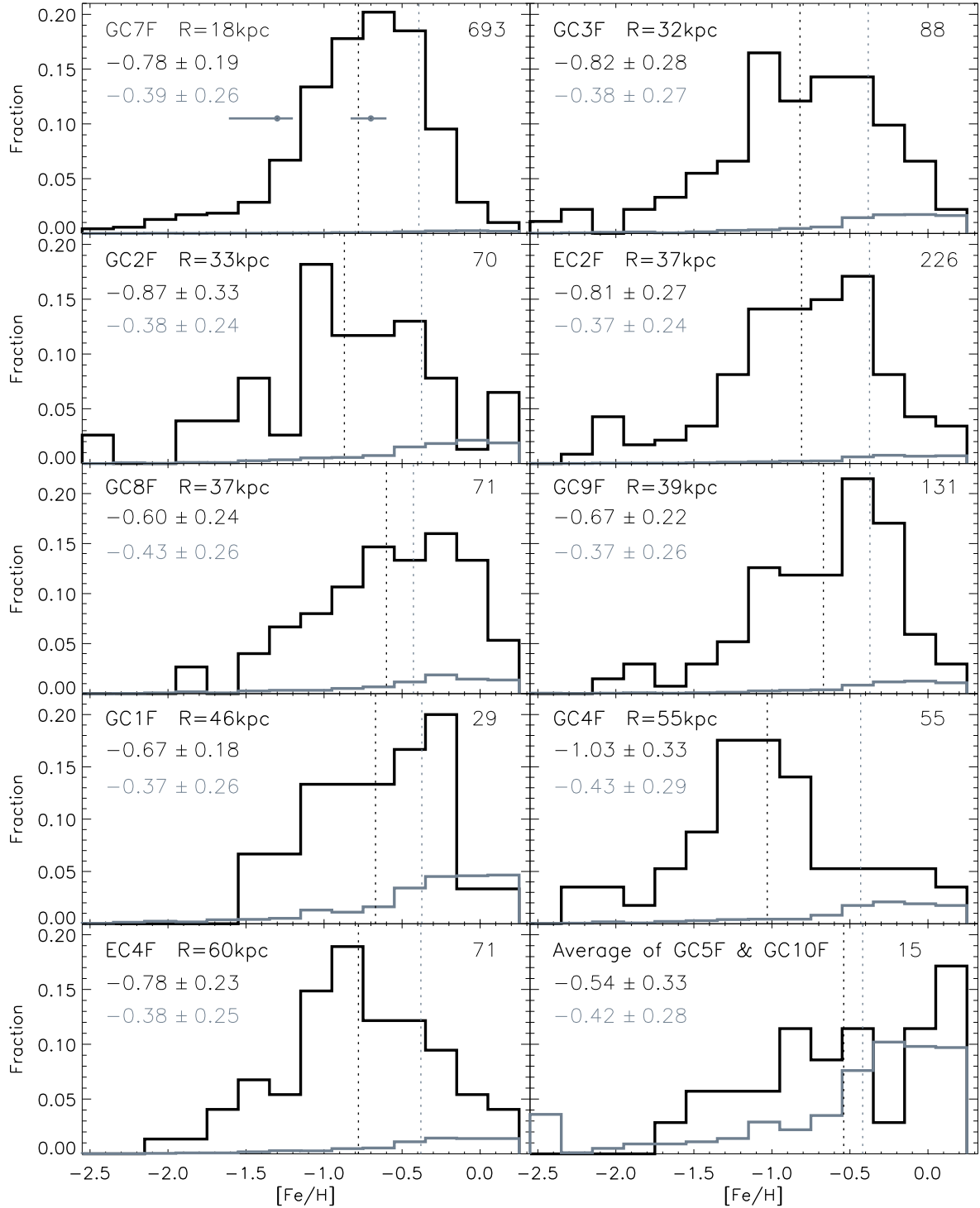}
  \caption{The MDFs for our outer halo fields assuming an age of 10
  Gyr and a Solar-scaled abundance distribution. Reported in each
  panel are the name and projected radial distance of the field plus
  the number of stars used to compute the MDF. The average metallicity
  $<\![\mbox{Fe/H}]\!>$ and its associated standard deviation, are
  listed below the field name and marked on the plot with a black
  dashed line. Gray lines and text indicate results for the foreground
  MW population.  The horizontal bars in the top left plot represent
  the typical error in the metallicity derived from the colour
  uncertainties at $m_{F814W,0} = 25.0$ mag. }
  \label{fig:mdfs}
\end{figure*}

In order to minimise contamination, the MDFs are computed from RGB
stars selected within the range $21.0<m_{F814W,0}<25.25$,
$m_{F606W,0}-m_{F814W,0}<1.7$ and sandwiched between the -2.5 and +0.3
dex isochrones. By selecting stars brighter than $m_{F814W,0} = 25.25$
mag, we limit the influence of unresolved background sources and
ensure a minimum completeness rate of $\approx 95\%$. This lower
magnitude limit marks the `knee' in the $m_{F814W,0}$ luminosity
function of the co-added fields beyond which the source count rises
very rapidly. The upper cut-off of $m_{F814W,0} = 21.0$ mag minimises
the inclusion of bright asymptotic giant branch (AGB) stars and the
colour cut ($m_{F606W,0} - m_{F814W,0} < 1.7$) avoids areas where the
density of stars on the CMD is very small and therefore liable to
increased influence from the MW foreground. An additional area ($23.9
< m_{F814W,0} < 24.6$) was cut to exclude the RC region. Although RGB
stars fainter than the RC have less metallicity sensitivity than their
brighter counterparts, they represent the bulk of the RGB populations
in these sparse fields and are essential for providing a sufficient
sample of stars. We have verified that our results do not change
significantly if we only use stars exclusively above or below the RC.

We present the resultant MDFs of the fields in Figure~\ref{fig:mdfs}
(solid black lines). The average metallicity and standard deviation of
each field are quoted in the panels and marked on the MDF by a black
dashed line (see also Table~1). Out to $R=60~{\rm kpc}$, the MDFs are
generally characterised by broad distributions which peak at
intermediate metallicities $<\![\mbox{Fe/H}]\!>\sim -0.6~ {\rm to}~
-1.0~{\rm dex}$, confirming inferences from visual inspection of the
CMDs alone. We reiterate that some of the metallicity spread might be
due to photometric errors and line-of-sight depth but these will not
dominate (see the top panel of Figure 4). Averaging over GC fields in
the 30--60~kpc range, the mean stellar metallicity is found to be
$<\![\mbox{Fe/H}]\!>\ = -0.78\pm0.14$. As previously mentioned, there
is no significant RGB detected in GC5F and GC10F and hence we are
unable to compute individual MDFs for these fields. However, they
provide a useful probe of the $`$contamination' in our MDF selection
box from non-M31 sources, namely unresolved background galaxies and
foreground dwarf stars. The bottom left-hand plot shows the MDF
derived for the average of the GC5F and GC10F fields.

Also shown in Figure~\ref{fig:mdfs} are solid grey lines representing
MDFs for the MW foreground towards each field as predicted by the
Besan\c{c}on Galactic model \citep{Besancon}. These are computed by
taking the simulations described previously and computing an MDF in an
identical manner to our GC fields. These are obviously not true
representations of the metallicities of MW foreground stars since this
population is dominated by dwarfs and not RGB stars on which our
analysis is based. Nevertheless, this procedure provides additional
insight into the degree to which the foreground might be biasing our
results. The average metallicity of the foreground is listed in grey
and highlighted with a grey dashed line. The foreground MDF peaks at
high metallicities ([Fe/H]$\approx -0.4$ dex) and has a long metal
poor tail, a shape which is rather different from that observed in any
of the fields out to 60~kpc. Thus, foreground stars alone cannot
explain the observed form of the MDFs with their strong peaks at
intermediate metallicities (although the foreground could explain a
large fraction of the Solar metallicity stars observed). We examined
the impact of subtracting the MW foreground MDFs from each field and
found it to be minimal. The average metallicities decreased by only
$\sim 0.02$ dex in all fields except for GC1F, which became 0.08 dex
more metal-poor, while the metallicity dispersions remained
essentially unchanged.

\begin{table}
\centering
  \caption{The effect of varying age and $[\alpha/Fe]$ on the derived
mean metallicity. \label{tab:shifts}}
  \begin{tabular}{@{} c c c  @{} }
\hline
 Age & $[\alpha/Fe]$ & $<[Fe/H]>_{shift}$ \\
  (Gyr) & (dex) & (dex)\\
\hline
12  &  0.0  &  -0.01 \\
10  &  0.0  &   0.00  \\
10  &  0.2  &  -0.04 \\
8   &  0.0  &  +0.02 \\
6   &  0.0  &  +0.05 \\
\hline
\end{tabular}

\end{table}

The shape of the average MDF of the GC5F$+$GC10F fields and the
Bescan\c{c}on model prediction are generally in good agreement which
suggests a large fraction of the contamination in the MDFs is due to
foreground stars. To test this further we performed a two-sided
Kolmogorov-Smirnov (KS) test and found a high probability (P = 0.95)
that the MDFs of GC5F$+$GC10F and the Besan\c{c}on foreground are
drawn from the same parent distribution. Note that this is true only
for bright stars which were incorporated into the MDFs. At magnitudes
fainter than $m_{F814W} = 25.25$ GC5F and GC10F suffer additional
contamination from unresolved background galaxies.  The surplus of
stars at [Fe/H]$< -0.8$ dex here may be a mixture of bright unresolved
background sources and true metal-poor M31 halo stars, or else it may
highlight a slight normalisation problem with the Galactic model.

Many of the MDFs contain stars with metallicities reaching as low as
$-2.5$ dex (this being the lower limit considered in our MDF
derivation). Interestingly, the fraction of metal-poor
($[\mbox{Fe/H}]\leq-1.3~{\rm dex}$) stars ($F_{mp}$) is relatively
invariant from field to field at $\sim 10-20\%$ although one field,
GC4F at $55~{\rm kpc}$, has a slightly larger metal-poor fraction (see
Table 1). Since it is possible that some of this population could be
due to AGB stars on the blue edge of the RGB, this number provides an
upper limit to the fraction of metal-poor halo stars sampled by our
fields. 

In deriving the MDFs, we have made the simplifying assumption that the
stellar populations are uniformly 10 Gyr old. The benefit of choosing
an old age to compute an MDF is that the position of an old star
within the RGB is mostly determined by its metallicity and largely
insensitive to its age. In reality, our fields will contain composite
stellar populations with varying age and metallicity
\citep[e.g.,][]{Brown08}. We have examined the effect of a straight
scaling of the age and alpha-enhancement of the isochrones on the
derived MDF (Table 2). Specifically, we reconstructed the MDF of each
field exactly as before but with a new set of isochrones as described
in Table~2. The mean shift in average metallicity
($<\![\mbox{Fe/H}]\!>_{shift}$) was calculated from the resulting MDFs
of all fields. When isochrones of 6 Gyr are used, there is only a
modest +0.05 dex increase in the average metallicity. On the other
hand, increasing $[\alpha/Fe]$ to +0.2 has the effect of shifting the
mean metallicity by -0.04 dex.  While the assumption of a fixed age
and alpha-enhancement will undoubtedly introduce uncertainties into
our results, we do not believe that these will be significant.

\section{Discussion}
\label{sec:disc}

Our analysis of the stellar populations in the vicinity of M31 outer
halo GCs indicates a field-to-field variation in the mean metallicity
of $<\![\mbox{Fe/H}]\!> \approx -0.6$ to $-1.0$~dex with a typical
dispersion of 0.2--0.3~dex. These relatively high metallicities are
supported by both CMD morphologies (e.g. mean RGB colours and the
presence of prominent RCs) as well as the MDFs. The average
metallicity over the range 30--60~kpc is [Fe/H]$=-0.8\pm 0.14$~dex,
with no evidence for a significant radial gradient.

Our findings for the M31 outer halo are in good agreement with
expectations from models in which stellar halos are built
predominantly from satellite accretion \citep{Font06halo, Font08,
deLucia08}. These models predict stellar halos to be highly structured
in their outskirts with discrete substructures becoming increasingly
dominant at large radius, an effect mostly driven by the long mixing
times in these parts \citep[e.g.,][]{Johnston08}. While the pencil
beam observations discussed here cannot test how the fraction of
substructure increases with radius, the field-to-field variations we
observe in RGB star counts provide good support for the notion that
the outer regions are poorly mixed in terms of stellar density.

\subsection{Connections to Known Substructures}

Even though our fields were selected purely on the basis of their
encompassing a luminous outer halo GC, Figure 1 reveals that a good
fraction of them lie on, or near, discrete substructures identified in
the CFHT MegaCam survey. This is intriguing but perhaps not completely
surprising given that a significant fraction of the halo luminosity
out to 60~kpc resides in such structures, which have a large covering
area on the sky (I07). GC7F, GC3F and GC2F are situated in reasonable
proximity to the diffuse Western shelf and GC9F is near to the
North-Eastern Shelf. Both of these substructures are most likely
comprised of debris stripped from the metal-rich giant stream
progenitor \citep{F07, Richardson08}. GC8F is located in close
proximity to the minor axis transverse Stream D, while EC4F probes
Stream C.  \cite{Chapman08} have recently found a bimodal
spectroscopic metallicity distribution for Stream C with peaks at
[Fe/H]=$-0.7$ and [Fe/H]=$-1.3$~dex.  Our MDF for EC4F also supports a
bimodal distribution with peaks at [Fe/H]=$-0.85$ and [Fe/H]=$-1.45$
dex, in good agreement. \cite{Fardal08} have shown it is plausible
that many of these outer halo features originate from the same event,
namely the accretion of the (metal-rich) giant stream
progenitor. Alternatively, even if the outer halo debris originates
from several different accretion events, the spatial coherence and
moderately high surface brightness of the streams imply that the
progenitors were of intermediate mass and hence metallicity
\citep{Font08}.

\begin{figure}
\centering 
\includegraphics[width=6.4cm,angle=90]{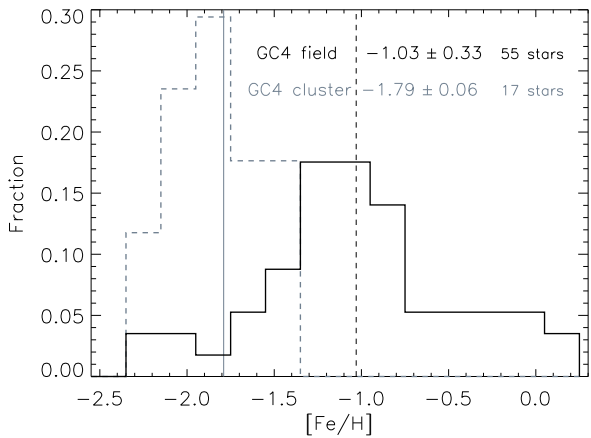}
  \caption{The relationship between the globular cluster GC4 and the
  associated field population. The solid black line represents the MDF
  of the field population (with the GC masked out to two $R_t$) while
  the grey dashed line is the globular cluster MDF measured up to
  three half-light radii ($R_h$). The average metallicity calculated
  over the whole metallicity range is listed beside the appropriate
  label and highlighted on the plot with a vertical line.}
  \label{fig:gc4}
\end{figure}

The most metal-poor population identified in our study is GC4F. This
field probes the `major axis diffuse structure' discovered by I07, who
find a metallicity of [Fe/H]=$-1.4$ dex from the colour of the RGB.
Our MDF shows a dominant peak at [Fe/H]$\sim -1.2$ dex in this
structure, which is in reasonable agreement. GC4 has been recently
shown to have a population of extra-tidal stars (\citealt{Federici07};
Tanvir et al. in prep.). In deriving the GC4F MDF we have taken
measures to avoid contamination by this population by masking out the
cluster to twice its tidal radius ($R_t = 22.4^{\prime\prime}$; Tanvir
et al. in prep.)  but it is worthwhile considering whether this has
been sufficient. In Figure 5 we compare the MDFs of the field and
cluster populations where the latter are defined as stars lying within
3 half-light radii (where $R_h = 1.3^{\prime\prime}$; Tanvir et al. in
prep.). It can be clearly seen that the cluster and field star MDFs
are very different. The cluster MDF shows a dominant peak at
[Fe/H]$\sim -1.85$~dex and an average metallicity of
$<\![\mbox{Fe/H}]\!> = -1.79 \pm 0.06$ dex. This is in excellent
agreement with the independently derived mean metallicities of
[Fe/H]=$-1.8$ to $-2.1$~dex reported in the literature
\citep{Galleti05, Galleti06, Mackey07}. The more enriched field
population ($<\!\mbox{[Fe/H]}\!> = -1.03 \pm 0.33$ dex) therefore
cannot be due to extra-tidal cluster stars, though some of the
metal-poor tail could be. This lends credence to the major axis
structure being a genuinely more metal-poor feature that is distinct
in origin from the giant stream debris.

\subsection{Radial Metallicity Trends}

\begin{figure}
\centering 
\includegraphics[width=8cm]{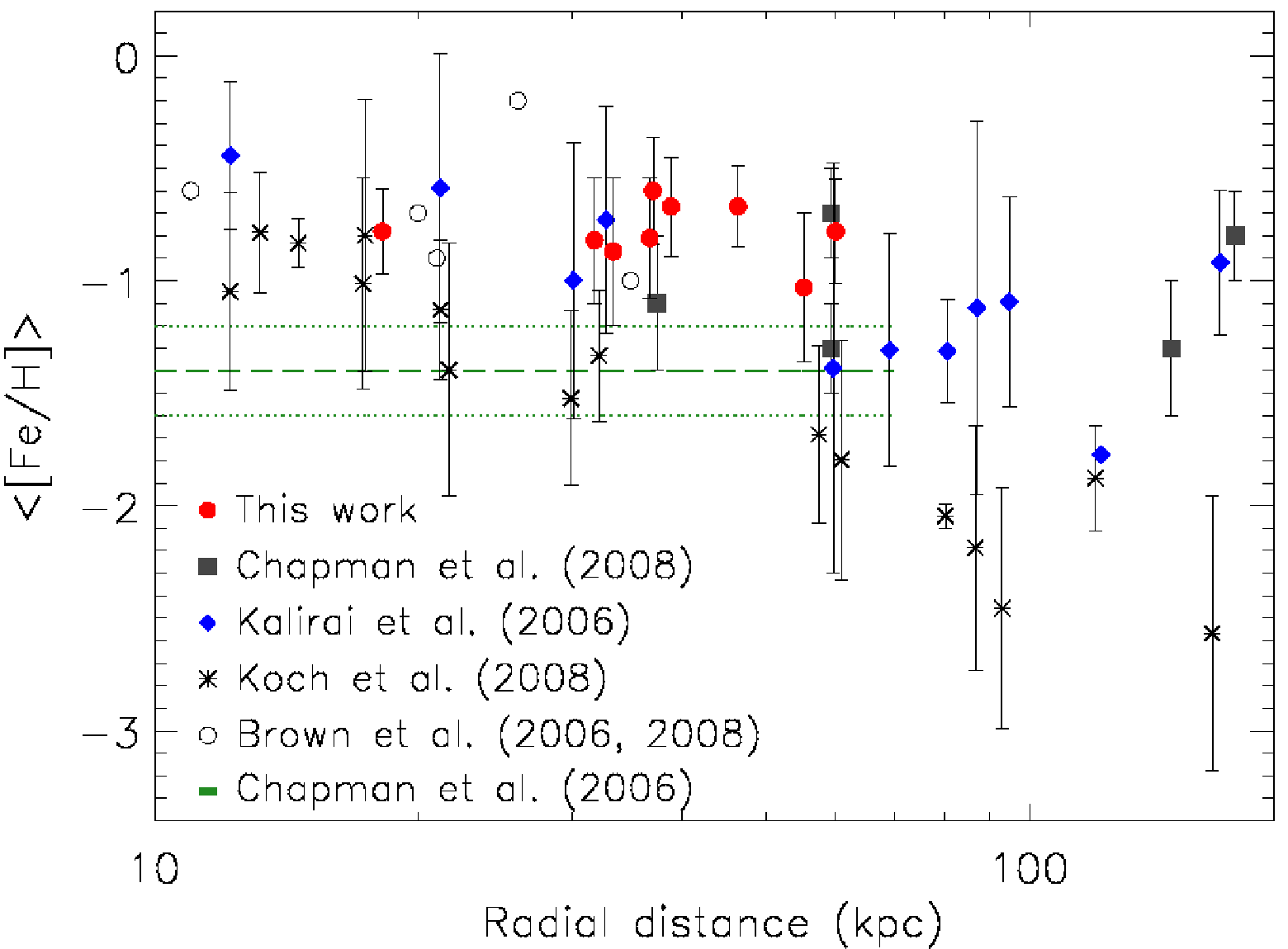}
  \caption{The mean metallicity and metallicity dispersion of the
  outer halo as a function of radius. Our results (red filled circles)
  are compared to all relevant studies in the literature. Gray squares
  represent the photometric [Fe/H] of stellar streams measured by
  \protect\cite{Chapman08}. Blue diamonds are the mean photometric
  metallicities of individual fields from \protect\cite{Kal06b} while the
  spectroscopic metallicities derived by \protect\cite{Koch08} are shown as
  black asterisks. The photometric metallicities of the
  \protect\cite{Brown06a, Brown08} fields are given by open circles. The green
  dashed (dotted) lines represent the spectroscopic metallicity
  (dispersion) traced by \protect\cite{Chapman06} between $10 < R ({\rm kpc})
  < 70$.}
  \label{fig:rad}
\end{figure}

In Figure~6 we present the radial metallicity distribution of M31's
outer halo as derived in this study and compare it to results in the
literature. We include here the spectroscopic studies of
\cite{Chapman08}, \cite{Kal06b}, \cite{Koch08} and \cite{Chapman06} as
well as the photometric studies by \cite{Brown06a, Brown08}. It is
important to note that all the spectroscopic studies invoke some
degree of kinematic selection to isolate the sample of halo stars
which will somewhat complicate the comparison of their results to
those from purely photometric work. Additionally, the metallicities
from \cite{Brown06a, Brown08} represent average values derived from
star formation history fits to the main sequence turn-off in
ultra-deep HST CMDs, a method rather different from that which we have
employed for our photometric metallicity derivation.

Figure~6 shows that the metallicities we infer in our fields agree
well with most other studies which have probed a similar radial range
although there is considerable dispersion at all radii. This
dispersion may be due to different methods of sample selection and
metallicity determination or to intrinsic metallicity variations in
stellar populations throughout the halo. In support of the last point
we highlight the $\sim$1 dex spread in metallicity at $R=60~{\rm kpc}$
where two separate streams (one metal-rich and one metal-poor - see
\citealt{Chapman08}) are probed.

Despite the general concordance between our results and those in the
literature, our measurements are systematically more metal-rich than
the spectroscopic metallicities measured by \cite{Chapman06} and
\cite{Koch08}. The discrepancy with Chapman et al. (2006) may be due
to the different sample selection; in their analysis, they windowed
out any stellar component that rotates within 100~km~s$^{-1}$ of the
thin disk over the radial range $10 < R ({\rm kpc}) < 70$. The
discrepancy with \cite{Koch08} is more puzzling. These authors have
re-analysed much of the same data as \cite{Kal06b} using a similar
halo sample selection method, but a different method of metallicity
determination, and find a strong metallicity gradient which reaches
[Fe/H]$= -1.8$ dex at $R\approx 60$~kpc. It is hard to reconcile such
a metal-poor population with the CMD morphology and MDF form that we
have observed in our outer halo GC fields.

Considered together, the results plotted in Figure~6 suggest that much
of the inner 60~kpc of M31's stellar halo is dominated by relatively
metal-rich populations ($-0.4 < [\mbox{Fe/H]} < -1.1$). This situation
appears vastly different from that in the Milky Way. \cite{Carollo08}
have recently derived metallicities for 1,235 blue horizontal branch stars in
the Galactic halo lying in the radial range 5--80~kpc and find the
population gets progressively more metal poor with radius, with the
dominant population at R$>30$~kpc having $<\![\mbox{Fe/H}]\!> \approx
-2$. While the halos of the Milky Way and M31 have been known for some
time to differ in their metallicities at smaller radii, this is the
first clear demonstration that these differences remain out to
significant radii. Unfortunately, few constraints exist on the
detailed nature of the M31 stellar populations beyond 80~kpc where
very metal-poor populations were hinted at by the ground-based
analysis of I07.  It may be that only in these parts of M31 a metal poor
stellar halo prevails.

Finally, it is worth considering the possibility that by examining the
fields surrounding luminous GCs at large radii we have introduced a
selection bias. The outer halo GCs targeted in the current study
comprise 11 out of the 40 ($\approx 25$\%) we have identified in this
particular quadrant \citep{Huxor08}.  These objects were chosen for
high resolution imagery with HST based based on the fact that they
were known at the time of proposal writing and span a broad range of
projected galactocentric radii; there are no signs that they are
unusual compared to other M31 halo GCs.  Within the Milky Way, at
least eight GCs have been possibly associated with the disrupting Sgr
dwarf galaxy, accounting for $\sim 20\%$ of all GCs at $R > 10~{\rm
  kpc}$ \citep{Bellazzini03b}. This raises the question as to whether
GCs at large galacto-centric radii are preferentially associated with
halo tidal stream debris, an intriguing idea which we will explore
more in future work.

 
\vspace{-0.5cm}

\section{Summary}

We have presented high precision photometry of the stellar populations
adjacent to 11 luminous outer halo GCs in M31 using HST/ACS data. Our
fields sample the extended halo in the radial range $18\la ~{\rmn{R}}
\la 100~{\rmn{kpc}}$ and reach a photometric depth of $\sim 2.5$
magnitudes below the horizontal branch (m$_{F814W,0}\sim 27$
mag). Eight pointings lie at $R > 35~{\rm kpc}$ and provide the first
detailed view of the properties of halo stars in these remote parts.

The colour-magnitude diagrams (CMDs) are well populated out to $\sim
60~{\rmn{kpc}}$ and exhibit relatively metal-rich red giant branches,
with the densest fields also showing evidence for prominent red
clumps. Using the Dartmouth isochrones, we have constructed MDFs which
confirm the presence of dominant populations with $<\![\mbox{Fe/H}]\!>
\approx -0.6$ to $-1.0$~dex and considerable internal metallicity
dispersions of 0.2 to 0.3 dex (assuming a 10~Gyr population and
scaled-Solar abundances). The average metallicity over the range
30--60~kpc is [Fe/H]$=-0.8\pm 0.14$~dex, with no evidence for a
significant radial gradient. Although some metal-poor stars
([Fe/H]$\leq -1.3$) are present in each field, they amount to
typically $\la 10-20\%$ of the total population, irrespective of
radius.

Significant inhomogeneities are present in the fields in terms of both
random fluctuations in star counts and field-to-field variations in
metallicity supporting the notion that the stellar populations are not
well-mixed beyond 30~kpc, in agreement with results from wide-field
surveys and predictions of models in which stellar halos assemble
predominantly from satellite accretion. Our photometric metallicity
measurements are in good agreement with most, but not all,
spectroscopic studies of M31's outer halo. Assuming that our GC fields
are unbiased probes of the dominant stellar populations in these
parts, we find that the M31 outer halo remains considerably more
metal-rich than that of the Milky Way out to at least 60~kpc.


\section*{Acknowledgements}

JCR acknowledges  the award of an  STFC studentship. AMNF,  ADM and AH
acknowledge  support from  a  Marie Curie  Excellence  Grant from  the
European Commission under contract MCEXT-CT-2005-025869. NRT
acknowledges financial support from a   STFC  Senior   Research
Fellowship. We thank Jay Gallagher, Alan McConnachie and the anonymous
referee for helpful suggestions  and Jasonjot Kalirai and Andreas Koch
for providing data for use in Figure~6.


\bibliographystyle{mn2e}
\bibliography{Richardson2009}

\bsp

\label{lastpage}

\end{document}